\documentclass{elsart}
\usepackage{epsf,amssymb}
\usepackage{graphicx}
\begin{document}
\begin{frontmatter}

\title{Long-range electronic interactions between adatoms 
on transition metal surfaces\thanksref{titlref}} 
\thanks[titlref]{This work was supported by the DFG (Schwerpunktprogramm 1153).}
\author[MPI]{V.S. Stepanyuk \corauthref{cor}},
\corauth[cor]{Corresponding author. E-mail: stepanyu@mpi-halle.de, 
                                    Phone: +49 345 5525429, 
                                    Fax: +49 345 5525446}
\author[MPI]{L. Niebergall},
\author[UNI]{A.N. Baranov},
\author[UNI]{W. Hergert},
\author[MPI]{P. Bruno}

\address[MPI]{Max-Planck-Institut f\"ur Mikrostrukturphysik, Weinberg
2, 06120 Halle, Germany }

\address[UNI]{Fachbereich Physik, Martin-Luther-Universit\"at,
Halle-Wittenberg, Friedemann-Bach-Platz 6,
 D-06099 Halle, Germany}

%\date{\today}
%\maketitle

\begin{abstract}
Ab initio calculations of  surface-state mediated interactions between 
Cu adatoms  on transition metal surfaces are presented. 
We concentrate on Co/Cu(111) and Co(0001) substrates and compare results 
with our calculations for Cu(111). Our studies show 
that surface states of Co/Cu(111) and Co(0001) are spin-polarized.   
We reveal  that long-range interactions between adatoms are mainly 
determined by 
$sp$-majority states. In contrast to Cu(111) and Co/Cu(111), the  
interaction between adatoms on Co(0001) 
is strongly suppressed at large adsorbate separations. 

\end{abstract}
\begin{keyword}
Ab initio calculations \sep spin-polarized surface states 
\sep long-range interactions

\PACS 
71.15.Mb \sep 71.20.Be \sep 73.20.At
\end{keyword}
\end{frontmatter}

\newpage

In the last few years  there has been a renewed interest in the study of 
indirect adsorbate interactions on metal surfaces predicted by Lau and 
Kohn in 1978\cite{c1}. This is at least in 
part due to the possibility to probe directly such interactions by 
scanning tunnelling microscopy (STM). A low temperature STM has 
allowed  to 
resolve  adsorbate interactions up to 80 \AA \cite{c2,c3}.  
It has been shown 
 that  oscillations of the electron density around  adsorbates 
in a two-dimensional (2D) electron system  can lead to a long-range, 
oscillatory, 
Friedel-type adsorbate-adsorbate interaction which decays 
as $1/d^2$. The required 2D nearly free electron gas could be realized in 
Shockley type surface states of metal surfaces.  For example,  
surface-state electrons  on the (111) surfaces of noble metals
form a  2D nearly free electron gas. 
Several  theoretical investigations have demonstrated that despite the fact that 
indirect adsorbate interactions are small (a few meV), they   can 
significantly influence the  growth of nanostructures\cite{c4,c5,c6}.

Very recent STM experiments and ab initio calculations have revealed 
that 
surface states of transition metal nanostructures   
can be spin-polarized\cite{c7,c8}. In particular, it has been shown that  
the 
electronic states of fcc Co 
monolayers and Co islands on Cu(111) are spin-polarized. Similar results 
have been 
reported for surface states of hcp Co(0001)\cite{c9}. In this paper, we 
present the first 
ab initio calculations of adsorbate-adsorbate interactions on transition 
metal  surfaces.
We concentrate on the interaction between Cu 
adatoms on the fcc Co/Cu(111) and the hcp Co(0001) substrates.  
Results are compared with the interaction on Cu(111).
We demonstrate that 
the substrate-mediated interactions on magnetic substrates  are 
mainly  determined  by $sp$-majority states.  
Our results show that 
adsorbate interactions on Co/Cu(111) are long-ranged and oscillatory. 
However, for Co(0001) we find that the substrate-mediated interactions are 
strongly suppressed at large distances. 
 
Our calculations are based on the density functional theory and
multiple-scattering approach  using the Korringa-Kohn-Rostoker
Green's function\linebreak method \cite{c6,c7,c10}. 
We treat the surface as an infinite two-dimensional perturbation of the
bulk.
Taking into account the 2D
periodicity of the ideal surface, we calculate the structural
Green's function by solving a Dyson equation self-consistently.
The consideration of adsorbate atoms on the surface destroyes the
translation symmetry. Therefore the Green's function of the
adsorbate adatom on the surface is calculated in a real
space formulation.
 The structural Green's function of the ideal surface  in real 
space representation is then
used as the reference Green's function for the calculation of the
adatom-surface system. 
Details of the method and
its first applications for calculations of 
surface-state electrons and  adsorbate-adsorbate interactions  
can be found in our previous work\cite{c6,c7,c10}.
Calculations
for the long-range interactions have been  performed for the relaxed and
unrelaxed positions of adatoms. 
However, we have found that the substrate-mediated interactions at
large distances are essentially unmodified by the inclusion of the
relaxation.

First, we present the results for the interaction between Cu adatoms 
on Cu(111).
Our calculation for the Cu(111) surface gives 
a surface-state Fermi wavelength $\lambda_F=29$ \AA\ and a surface-state 
band edge at E$_0$=-0.5 eV below the Fermi level.
Due to the long Fermi-wavelength the confinement property of the
Cu adatom should exist for large distances around the Cu adsorbate.
The scattering of surface state electrons by Cu adatoms leads to
 quantum interference patterns  and to the 
long-range interactions between the two Cu adatoms\cite{c6}.  
Our 
calculations for the interaction energy between Cu adatoms on Cu(111) are
presented in Fig.1.\\

\begin{figure}[htp]
\begin{center}
   \includegraphics[width=10cm]{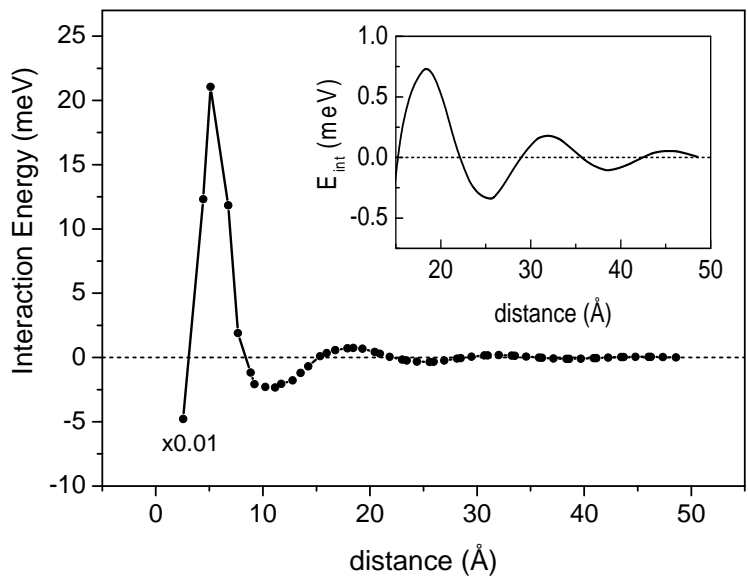}
\caption {Calculated interaction energies between two Cu adatoms on
Cu(111). Inset shows the long-range oscillatory interaction at large
distances.}\label{fig1}
\end{center}
\end{figure} 

These results
show that the interaction energy is oscillatory with a period of
about 15 \AA. The envelope of the magnitude decays as $1/d^2$ in
agreement with the prediction of Lau and Kohn\cite{c1}. 

In recent STM experiments on large Co islands supported on  Cu(111) the 
standing wave 
patterns in the local density of states (LDOS) due to the quantum 
interference of surface-state electrons have been observed\cite{c7,c8}. Ab 
initio 
calculations\cite{c7} have revealed that a majority free-electron like 
surface 
states  give rise to LDOS oscillations on Co islands. 
In Fig.2, as an example we show the contour plot for majority and minority 
spectral densities of two monolayers of Co on Cu(111).

\begin{figure}[htp]
\begin{center}
 \includegraphics[width=14cm]{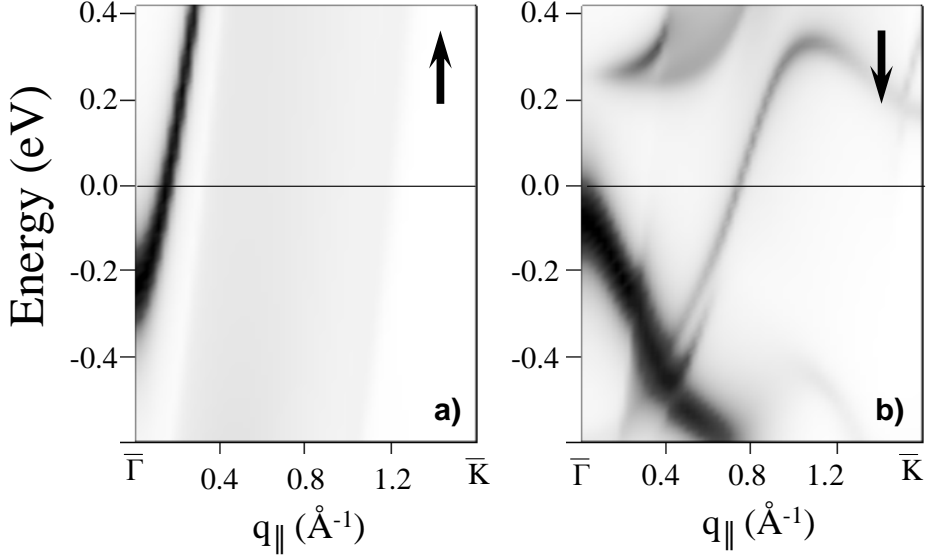}
\caption{Contour plot for majority(a) and minority(b) spectral densities
of surface-state electrons on the fcc  2Co/Cu(111); the Fermi-energy is
choosen as zero.}
\end{center}
\end{figure}

We find that 
the majority $sp$  
states have a parabolic dispersion relation, described by an onset  
below the Fermi level at $E_0$=0.28~eV and an effective mass $m^*=0.50m_e$. 
Compared to Cu(111), a surface-state Fermi wavelength for 2Co/Cu(111)
is found to be significantly larger ($\lambda_F=48$~\AA).
In the minority channel  the $d$ contribution is found to be dominant. The 
strong  feature below the Fermi level (cf. Fig.2b) is mainly 
determined by 
the $d$-states with a small $sp$ contribution. These states lead to a 
strong localized peak at about 
0.4 eV below the Fermi energy observed in experiments\cite{c7,c8}. 

Our calculations show that the LDOS around a single Cu adatom on\linebreak 
2Co/Cu(001) displays the long-range Friedel oscillations  
caused by the quantum interference of the $sp$-majority surface-state 
electrons. These oscillations  lead to a long-range  
interaction between adatoms. Results presented  in Fig.3.

\begin{figure}[htp]
\begin{center}
\includegraphics[width=10cm]{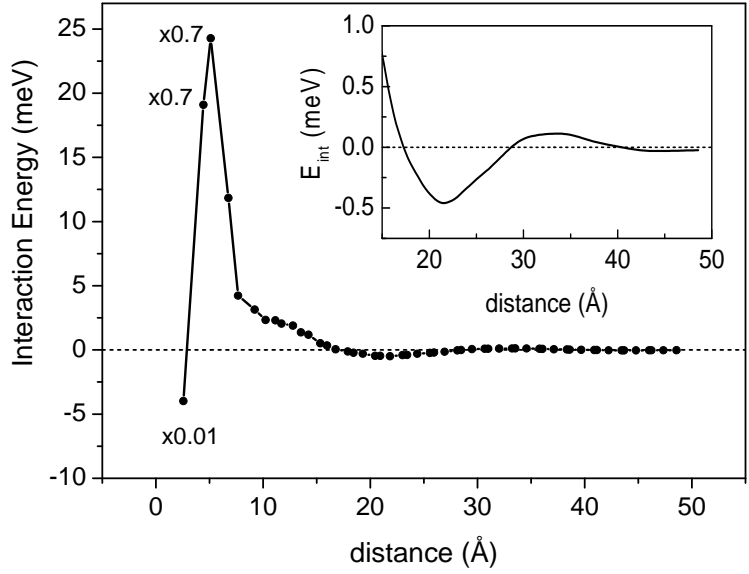}
\caption{Calculated interaction energies between two Cu adatoms on the fcc
2Co/Cu(111). Inset shows the long-range oscillatory interaction at large
distances.}
\end{center}
\end{figure}
 
clearly demonstrate that the interaction between Cu adatoms is 
oscillatory with a period of about 24 \AA\ corresponding to  the Fermi 
wavelength of the majority $sp$ states of fcc Co (cf. Fig.2a).   
It is important to note that our studies for different number of Co layers 
on Cu(111) have shown that the surface-state band edge of the majority 
electrons increases with coverage\cite{c7}. 
Therefore, there is the possibility of 
tailoring  the long-range interactions on Co/Cu(111) by  variation of the
coverage. Our results for different number of Co monolayers on Cu(111) 
will be presented elsewhere.

Finally, we  discuss our results for the hcp Co(0001) substrate.
Recent spin-polarized STM experiments of Okuno et al.,\cite{c9} have 
found a spin-polarized surface state at -0.43 eV relative to the Fermi energy. 
Our calculations of the spectral densities for majority and minority
states are shown in Fig.4.

\begin{figure}[htp]
\begin{center}
\includegraphics[width=14cm]{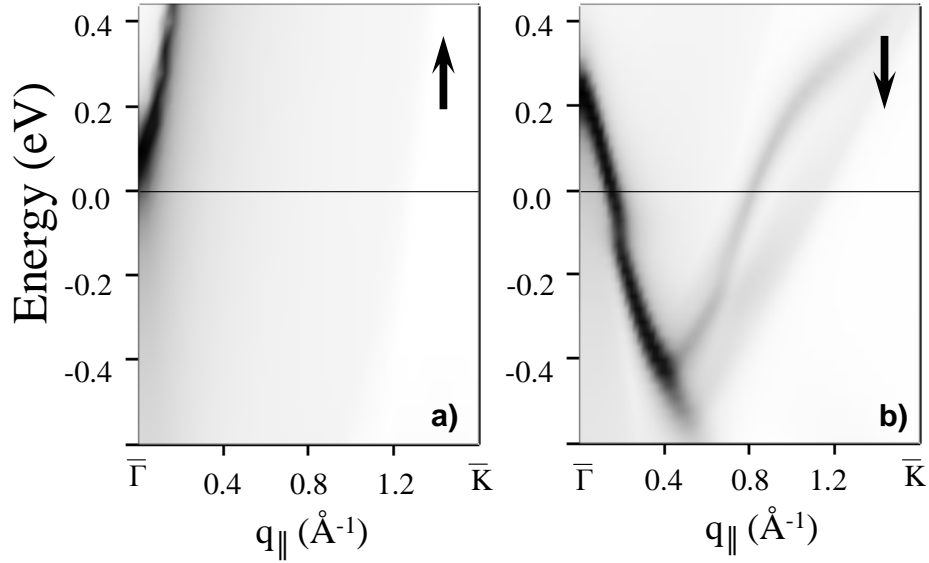}
\caption{ Contour plot for majority(a) and minority(b) spectral densities
of surface-state electrons on the hcp Co(0001). }
\end{center}
\end{figure}\label{fig4}

One can see that a minority surface state
exists and the energy of its minimum  
agrees well with the experiment\cite{c9}.  
Our analysis shows that this state  has $d$ character similar 
to the minority state of fcc Co (cf. Fig.2b). 
However, in contrast to the fcc Co, the $sp$-majority states of the hcp 
Co(0001) 
are  strongly shifted to higher energies and become  unoccupied. 
These results suggest that the substrate-mediated interaction between 
adatoms on Co(0001) could be very much different from that 
on Cu(111) and fcc Co substrates.
Our calculations presented in Fig.5 show that the interaction energy 
between Cu adatoms on Co(0001) decays very fast and practically vanishes 
for adatom-adatom separation larger than 15 \AA .

\begin{figure}[htp]
\begin{center}
\includegraphics[width=10cm]{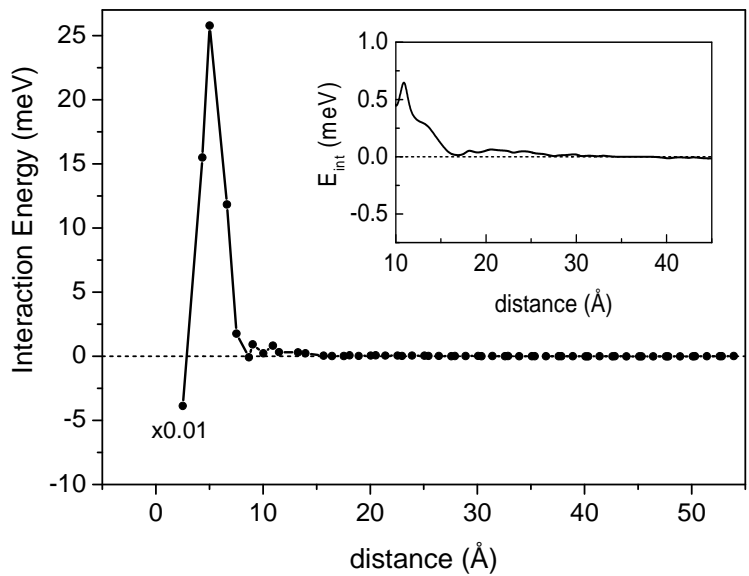}
\caption{Calculated interaction energies between two Cu adatoms on the hcp
Co(0001). Inset shows the interaction at large distances. }\label{fig5}
\end{center}
\end{figure} 

In summary, we  have performed first  ab initio
calculations for the substrate mediated  adsorbate-adsorbate interactions
on transition metal surfaces. 
We have found that the surface states on the fcc Co and the hcp Co 
substrates are 
spin-polarized. Our results predict that the spin-polarization of 
surface-state electrons can strongly affect interactions between adatoms.
We  reveal that mainly majority $sp$-states determine the 
interaction at large adatom-adatom separations.

\end{document}